\documentstyle{article}

\def\be{\begin{equation}}
\def\ee{\end{equation}}
\def\bd{\begin{displaymath}}
\def\ed{\end{displaymath}}

\def\be{\begin{equation}}
\def\ee{\end{equation}}

\def\bm{\begin{displaymath}}
\def\ebm{\end{displaymath}}

\begin{document}

\author{{\large S.G.Rubin}\thanks{e-mail: serg.rubin@mtu-net.ru}}
\title{{\bf BAND STRUCTURE OF LOCAL PAIRS. \\
MATHEMATHICAL TOOLS.}}
\date{Moscow Engineering Physics Institute (Technical University), Moscow, Russia}
\maketitle

\begin{abstract}
A system of strongly interacting fermions in a solid state is discussed. As
an example, a structure of singlet and triplet coupled 2-particle states and
their excitation spectra are investigated. It is shown that an account of
intersite fermion interaction leads to new boson modes. Their number depends
on dimensionality of a system. It is shown also that a reasonable choice of
Hamiltonian parameters leads to a small effective mass of local pairs. A
problem of interaction of coupled states with external fields are discussed.
The calculations are carried out for intermediate and strong fermion --
fermion coupling. The same analysis can be performed for another sorts of
pairings like d-wave.
\end{abstract}


\bigskip

\section{\bf Introduction}

\bigskip The discovery of high-temperature superconductors (HTSC) that
differ drastically from conventional superconductors leads to a great number
of theoretical models. A small coherence length in HTSC--materials indicates
that a size of fermion pairs, which is crucial for the superconducting
properties, should be of the order of lattice parameter.

In this paper the formalism of tightly binding fermions in solids is
developed. The main tool is extended Hubbard model with a fermion-fermion
intersite attraction. The origin of the attraction could be 'phononic',
'excitonic' \cite{Little}, 'plasmonic' \cite{Frohlich} or 'magnonic' \cite
{Schri}.

Last years indicate the power of the bipolaron theory \cite{Krebs}, \cite
{Alex} and one could identify the local pairs and bipolarons -- local pairs
with phonon mechanism of 2-fermion attraction. The presence of the
attraction changes the band structure significantly. Fermions being placed
on different sites are coupled into pairs. The continuum of electron-hole
pairs is situated above the local pair bands and does not overlap with them
both in the normal and superconducting phases -- see discussion in \cite
{Sofo}. It is the main difference of BCS-like theories and local pair ones.

The pair size seems to be of the same order of the distance between pairs
that appears to be rather difficult for theoretical analysis. Nevertheless
there are some papers ~\cite{Sofo}, ~\cite{drechsler} that are devoted to
this complex intermediate region of crossover from Cooper pair regime to
local pair one.

A presence of a superconducting state indicates an existence of boson type
excitations, that are able to form a condensate at low temperatures. But
there are fermions (electrons) in solids as the only charge carriers from
the beginning that interact with phonons and other quasiparticles in
crystal. This interaction in HTSC -- materials is very complicated both for
experimental and theoretical investigations. As a result, a large variety of
superconductivity models has been developed. Most of papers deal with a
ground state and excitation spectra on the basis of an attractive or
repulsive Hubbard model (see review ~\cite{micnas}). Although a
generalization to a more realistic and complicated situation is not obvious,
several interesting papers have been recently appeared ~\cite{Sofo}, ~\cite
{boer}, ~\cite{Verdozzi}. In ~\cite{boer} the authors considered in detail a
ground state of such a system with rather general form of interaction in
site representation. In ~\cite{Verdozzi} the Green function method was
applied to the boson spectra excitations. But the latter work was motivated
by Auger experiments in noble metals and only fermion -- fermion repulsion
was taken into account.

This work is devoted to the investigation of the boson excitations - their
band structure, interaction with each other and with external fields.

We start with rather general form of fermion -- fermion interaction to be
able to describe a large variety of interaction forms in different HTSC --
materials. To be more definite, we postulate the effective Hamiltonian for
fermi excitations in site representation in the form: 
\begin{eqnarray}
H=-\mu \sum_M c_M^{+}c_M+\!\sum_{M,N}t({\bf m}-{\bf n})c_M^{+}c_N+  \nonumber
\\
\frac{1}{4}\sum_{MNM^{\prime }N^{\prime }}U_{MN,M^{\prime }N^{\prime
}}c_M^{+}c_N^{+}c_{M^{\prime }}c_{N^{\prime }}+\sum_{MN}B_{MN}c_M^{+}c_N.
\label{e1}
\end{eqnarray}
Here ${\mu }$ is the chemical potential, and for example $M=({\bf m},s)$,
where ${\bf m}$ represents the site index, $s$ is the spin projection of the
fermion, $c_M^{+},c_M$ are the operators for fermion creation and
annihilation on the site ${\bf m}$ with the spin projection $s=\pm \frac{1}{2%
}$, $U_{MN,M^{\prime }N^{\prime }}$ - is the matrix element of fermion
interaction. This matrix is antisymmetric in its first and second pairs of
indices separately: $U_{MN,M^{\prime }N^{\prime }}=-U_{NM,M^{\prime
}N^{\prime }}=-U_{MN,N^{\prime }M^{\prime }}.$ A hopping integral $t({\bf m}-%
{\bf n})$ characterizes the fermion motion and describes a kinetic energy in
a momentum representation. $B_{MN}$ represents an arbitrary external field.

It must be stressed here that only fermi statistics of primordial
quasiparticles is important. One can keep in mind polarons, holes, magnons
and so on. The question is if the interaction is strong enough to form the
local pairs or we have Cooper pairs in high-$T_c$ superconductors as in the
ordinary ones. A mathematical method in the latter case has been developed
quite accurately. The method, developed in this paper for the description of
the local pairs is based on a functional integral technique ~\cite{b4}, that
was widely used in different theoretical investigations (see, for example, ~
\cite{b2}, ~\cite{RubKon}, ~\cite{cross}). It was applied to a
single-polaron problem ~\cite{polaron} and is used here for the local pair
description of strongly interacting fermions. Calculations appear to be
rather simple in the case of short - distance interaction. Furthermore
mathematical difficulties increase slightly with a number of accounted sites.

Below the general formalism is presented for an arbitrary form of fermion --
fermion interaction. Detailed calculations of spectra excitations and a
structure of the pairs are discussed. Several applications of the developed
method were considered in papers \cite{RubMel}, devoted to tunneling and 
\cite{RubTer}, where it was shown that experimental data on neutron
scattering and infrared absorption spectra are in good agreement with
predictions of local pair model.

\bigskip

\section{\bf General formalism}

\bigskip

\subsection{\bf Main equations}

The Hamiltonian $H$ in the form (\ref{e1}) is quite general and its origin
is not discussed here. The external field $B$ is written in a general form
too. This form includes the interaction with electromagnetic field and
neutrons, for example. Let's start with partition function of the system in
a functional integral form ~\cite{b2}:

\begin{eqnarray}
&&Z=Spe^{-\beta H}=\int Dc^{*}Dc\exp {S[c^{*},c]},  \nonumber \\
&&S[c^{*},c]=\int_{0}^{\beta} d\tau [\sum_{A} {c^*}_A \dot{c}_A - H(c^{*},c)]
\nonumber \\
&&=\int_0^\beta d\tau [\sum _A c_A ^{*}\dot{c}_A + \sum _{AB}r_{AB}c_A
^{*}c_B - \sum _{ABA^{\prime }B^{\prime }}\frac{1}{4}U_{AB;A^{\prime
}B^{\prime }}c_A ^{*}c_B ^{*}c_{A^{\prime }}c_{B^{\prime }}],  \label{e2}
\end{eqnarray}
\[
r_{AB}=(\mu -\omega _{p_A})\delta _{AB}.
\]
Expression (\ref{e2}) is written in momentum representation where complex
index $A=(p_A ,s_A )$ includes quasimomentum and spin projection of a
fermion; $\omega _{p_A}$ is the fermion spectrum determined as a Fourier
transform of the hopping integral; $c_A^{*}(\tau ),c_A(\tau )$ are
independent grassman variables; $\beta =1/T$ is the inverse temperature. The
external field $B$ is omitted in this section. To get rid of the four
grassman variables product in the exponent and introduce the bosonic
variables $\psi _{AB}(\tau )$ one can use the Hubbard -- Stratonovich
transformation ~\cite{HS}, \cite{drechsler} which is usually applied to
consideration of bosonic fields of one argument that describes boson motion
as a whole. In this latter case all the information about the internal
structure of the bound state is lost, though the results of conventional BCS
-- theory of superconductivity are reproduced ~\cite{b2},~\cite{Svidz}. But
as far as we intend to study the internal pair structure and the interaction
of the pairs with external fields we should use more complex bosonic fields $%
\psi _{AB}(\tau ),\psi _{AB}(\tau )^{*}.$ In this case expression (\ref{e2})
is transformed into:

\begin{eqnarray}
&&Z=\int Dc^{*}DcD\psi ^{*}D\psi e^{S_e},  \nonumber \\
&&S_{e}=\int _0 ^{\beta} d\tau [\sum _A c_A ^{*}\dot{c}_A + \sum _{AB}\left[
r_{AB}c_A^{*}c_B +\frac{1}{2}\psi _{AB}^{*}c_A c_B +\frac{1}{2}\psi _{AB}c_A
^{*}c_B ^{*}\right]  \nonumber \\
&&+\sum _{ABA^{\prime }B^{\prime }}\psi _{AB}^{*}U_{AB;A^{\prime }B^{\prime
}}^{-1}\psi _{A^{\prime }B^{\prime }}].  \label{e3}
\end{eqnarray}

Our central problem is to obtain equations for the wave functions of the
coupled states and to find their solutions for the chosen form of
interaction term. An imaginary ''time'' Fourier -- transformation gives:

\begin{eqnarray*}
S_e=\sum _{\varepsilon AB}(r_{AB}+i\varepsilon \delta
_{AB})c_A^{*}(\varepsilon )c_B(\varepsilon )+\frac 12\sum _{\varepsilon
\varepsilon ^{\prime }AB}\psi _{AB}^{*}(\varepsilon +\varepsilon ^{\prime
})c_A(\varepsilon )c_B(\varepsilon ^{\prime }) \\
+\frac{1}{2}\sum _{\varepsilon \varepsilon ^{\prime }AB}\psi
_{AB}(\varepsilon +\varepsilon ^{\prime })c_A^{*}(\varepsilon
)c_B^{*}(\varepsilon ^{\prime })+\sum _{EABA^{\prime }B^{\prime }}\psi
_{AB}^{*}(E)U_{AB;A^{\prime }B^{\prime }}^{-1}\psi _{A^{\prime }B^{\prime
}}(E),
\end{eqnarray*}
where Matsubara frequencies $\varepsilon =(2n+1)\pi /\beta , \varepsilon
^{\prime }=(2n^{\prime }+1)\pi /\beta $ due to antiperiodic boundary
conditions for fermionic variables and $E=2\pi n/\beta $ due to periodic
boundary conditions for boson dynamic variables. Now one can integrate out
the grassman variables \cite{b2}:

\begin{eqnarray}
&&Z=\int D\psi ^{*}D\psi e^{S_b} ,  \nonumber \\
&&S_b =\sum _{ABA^{\prime }B^{\prime }E}\psi _{AB} ^{*}(E) U_{AB;A^{\prime
}B^{\prime }} ^{-1}\psi _{A^{\prime }B^{\prime }}(E)+ Tr(lnY),  \label{e4}
\end{eqnarray}

\begin{eqnarray*}
Y_{AB}(\varepsilon ,\varepsilon ^{\prime})=\left( 
\begin{array}{cc}
(r_{AB} +i\varepsilon \delta _{AB})\delta _{\varepsilon ,\varepsilon
^{\prime}} & \psi_{AB}( \varepsilon + \varepsilon ^{\prime}) \\ 
\psi^* _{AB}(\varepsilon + \varepsilon ^{\prime}) & -(r_{AB} +i\varepsilon
\delta _{AB})\delta _{\varepsilon ,\varepsilon ^{\prime}}
\end{array}
\right).
\end{eqnarray*}

Boson dynamic variables $\psi _{AB}^{*},\psi _{AB}$ having been introduced,
the expression (\ref{e4}) remains not very suitable for the further
analysis. Next step is to expand the action $S_b$ around a classical action $%
S_b [\psi _{cl}]$. ($\delta S_b/\delta \psi _{cl}=0$ by definition). The
simplest case $\psi _{cl}=0$, takes place: a) in a normal phase (by
definition), b) near critical temperature, where a condensate density is
small, and c) in a low fermion density limit. Calculations for $\psi
_{cl}\neq 0$ were discussed in \cite{RubKon}. Here we will be interested in
the normal phase of the system.

Let's expand the effective action $S_b$ (\ref{e4}) in Tailor series up to
the second order in fields $\psi _{AB}^{*},\psi _{AB}$ around trivial
classical trajectory $\psi _{cl}=0$. A fourth order in the fluctuation
describes interaction of pairs and was discussed in ~\cite{b1}. As it was
shown in ref. ~\cite{zetp}, this term is small in a low-density region
and/or for a weak fermion interaction. Here we neglect the pair interaction
and limit ourselves to the study of boson excitation spectra and their
interaction with external fields. Simple, but tedious calculations lead to
quadratic part of the effective boson action:

\begin{eqnarray*}
&S_b^{(2)}=\sum _{ABA^{\prime }B^{\prime }E}\psi _{AB}^{*}(E)\tilde
D_{AB,A^{\prime }B^{\prime }}(E)\psi _{A^{\prime }B^{\prime }}(E), \\
\\
&\tilde{D}_{AB,A^{\prime}B^{\prime}}(E) =\frac{1}{4} \cdot {\frac{\strut {%
th(\beta \Omega _A /2) +th(\beta \Omega _B /2)} }{\displaystyle {-iE+\Omega
_A + \Omega _B}}} \delta _{AB^{\prime}}\delta _{A^{\prime}B} + U^{
-1}_{A^{\prime}B^{\prime};AB}\; , \\
\\
&\Omega _A =\omega _{{\bf p} _A }-\mu.
\end{eqnarray*}

To remove the denominator in the first term and the inverse power of the
interaction operator U, one has to make the following substitution of
variables in the functional integral (\ref{e4}):{\rm \ $U^{-1}\psi
\rightarrow \psi ,\psi _{AB}^{*}\cdot (iE\Omega _A-\Omega _B)\rightarrow
\psi _{AB}^{*}$, }. This procedure leads to the partition function in the
form:

\begin{eqnarray}
Z=\int D\psi ^* D\psi \exp \{ \sum _{ABA^{\prime}B^{\prime}E} \psi ^*
_{AB}(-D_{AB,A^{\prime}B^{\prime}} +iE\delta _{AA^{\prime}}\delta
_{BB^{\prime}})\psi _{A^{\prime}B^{\prime}}\},  \label{e5}
\end{eqnarray}
where 
\begin{eqnarray}
D_{AB,A^{\prime}B^{\prime}}=(\Omega _A +\Omega _B)\delta
_{AA^{\prime}}\delta _{BB^{\prime}} + W_{AB,B^{\prime}A^{\prime}},
\label{n1}
\end{eqnarray}

\begin{equation}
W_{AB;B^{\prime}A^{\prime}}=\frac{1}{4} \cdot (th\frac{\beta \Omega _A}{2}
+th\frac{\beta \Omega _B}{2})U_{AB;B^{\prime}A^{\prime}}.
\end{equation}
Now we have to diagonalize the quadratic form of the exponent, that is to
solve the equation:

\begin{eqnarray}
\sum _{A^{\prime }B^{\prime }} D_{AB;A^{\prime }B^{\prime }}u_{A^{\prime
}B^{\prime }}^{(n)}=\lambda _nu_{AB}^{(n)}.  \label{e6}
\end{eqnarray}
Complex index $n$ numerating the eigenstates of the system will be shown to
include quasimomentum of the local pair, its spin and a band number. The
eigenvalues $\lambda _n$ obey the secular equation: 
\begin{eqnarray}  \label{e7}
Det(D_{AB;A^{\prime }B^{\prime }}-\lambda _n \delta _{AA^{\prime }}\delta
_{BB^{\prime }})=0.
\end{eqnarray}
The eigenfunctions are chosen to be orthonormal: 
\begin{eqnarray}
\sum _{AB}u_{AB}^{(n)*}u_{AB}^{(n^{\prime })}=\delta _{nn^{\prime }}.
\label{norm.u}
\end{eqnarray}
The final step consists of introducing new variables $C_n(E)$ in the
integral (\ref{e5}): 
\begin{eqnarray*}
\psi _{AB}=\sum _nC_n(E)u_{AB}^{(n)};\ \psi _{AB}^{*}=\sum
_nC_n^{*}(E)u_{AB}^{(n)*}.
\end{eqnarray*}
This leads to the expression: 
\begin{eqnarray}
Z=\prod_n\int DC_n^{*}DC_n\exp \{C_n^{*}(iE-\lambda _n)C_n\}.  \label{e8}
\end{eqnarray}
This form of the partition function clarifies the meaning of the new
variables $C_n^{*},C_n$ as the wave functions of bosonic excitations of
energy $\lambda _n$. It will be shown below that $u_{AB}^{(n)}$ functions
describing the internal structure of the bound state must be taken into
account in the analysis of the pair interaction with an external field. Eq.(%
\ref{e7}) is the only one we need to calculate bosonic -- type spectra
excitations. The pair energy depends on its quasimomentum and spin
projection. Quasimomentum and spin conservation can be accounted for in the
form of interaction term: 
\begin{eqnarray}  \label{n2}
W_{AB,B^{\prime }A^{\prime }}=\frac{1}{4} w_{AB,B^{\prime }A^{\prime
}}\delta (A+B-A^{\prime }-B^{\prime })
\end{eqnarray}
Index $A=({\bf p}_A,s_A)$ includes both the momentum ${\bf p}_A$ and the
spin projection $s_A$ of the fermion. Let variables $R=A+B,$ $R^{\prime
}=A^{\prime }+B^{\prime }$ characterize the momentum and spin projection of
the coupled state as a whole. One can easily find from Eqs.(\ref{n1},\ref{n2}%
), that matrix $D$ is diagonal in indices $R,R^{\prime }$. Several energy
bands may exist and we specify them by additional index $\alpha $. It means
that the complex index $n$ in the expression (\ref{e8}) should incorporate
the quasimomentum ${\bf p}_K$ of the 2-particle state as well as its spin
projection $\sigma _K=0,\pm 1,$ united in the common index $K=({\bf p}%
_K,\sigma _K),$ and eigenfunctions of the matrix $D$ should have the form: 
\begin{eqnarray}  \label{setx}
u_{AB}^{(n)}\equiv v_{RA}^{(n)}=v_{RA}^{(\kappa ,\alpha ,K)}=\delta
_{K,R}x_{KA}^{(\kappa ,\alpha )},
\end{eqnarray}
where $\kappa =0(1)$ corresponds to singlet (triplet) case. The substitution
of expression (\ref{setx}) into Eq.(\ref{e6}) along with the normalization
condition term (\ref{norm.u}) lead to:

\begin{eqnarray}  \label{e9}
[\Omega _A+\Omega _{K-A}-\lambda _\alpha (K)]x_{KA}^{(\kappa ,\alpha )}+%
\frac{1}{4}\sum _{A^{\prime }}w_{A,K-A;K- A^{\prime },A^{\prime
}}x_{KA^{\prime }}^{(\kappa ,\alpha )}=0,
\end{eqnarray}
and 
\begin{eqnarray}  \label{norm.x}
\sum _Ax_{KA}^{(\kappa ,\alpha )*}x_{KA}^{(\kappa ^{\prime },\alpha ^{\prime
})}=\delta _{\kappa \kappa ^{\prime }}\delta _{\alpha \alpha ^{\prime }}.
\end{eqnarray}

These equations determine the pair structure $x_{KA}^{(\kappa ,\alpha )}$
and the excitation spectra $\lambda _{\kappa ,\alpha }(K)$ The previous
analysis shows that for $\kappa =0$ the solution is antisymmetrical in the
fermion spin indices ($x^{(0,\alpha )}_{K,A}=-x^{(0,\alpha )}_{K,K-A}$)
while it is symmetrical for $\kappa =1$. It confirms the notation $\kappa =0$
for the singlet pairs and $\kappa=1$ for the triplet ones.

\bigskip

\subsection{\bf Equations for triplet and singlet states}

\bigskip

Cooper pairs in the BCS theory are singlets. There are many reasons to think
that in HTSC -- materials triplet pairs exist along with singlet ones \cite
{triplet}. Both of these possibilities are included in the system of
equations (\ref{e9}) and the task is to write down the equations for the
singlet and triplet states separately. It is convenient to redefine the wave
functions of the coupled states and the interaction term $w$ by extracting
the spin variables, that are included in the indices $A$, $A^{\prime }$ and
so on: 
\begin{eqnarray*}
&x_{\sigma ,s}^{(\kappa ,\alpha )}({\bf k},{\bf p})\equiv x_{KA}^{(\kappa
,\alpha )}, \\
&w_{s,\sigma -s,\sigma -s^{\prime },s^{\prime }}\left( {\bf p},{\bf k} -{\bf %
p} ;{\bf k}-{\bf p}^{\prime },{\bf p}^{\prime }\right) \equiv
w_{A,K-A;K-A^{\prime },A^{\prime }} \\
\end{eqnarray*}
Here as usual: $K=({\bf k},\sigma );A=({\bf p},s);A=({\bf p^{\prime }}
,s^{\prime });\sigma =0,\pm 1;s=\pm \frac{1}{2}$. Lower (spin) indices of
the interaction $w$ are restricted to the values $\pm \frac{1}{2}$ by
definition.

Let us rewrite Eq.(\ref{e9}) using these notations for each spin projection $%
\sigma $ of a coupled state. The equation for $\sigma =1$ becomes:

\begin{eqnarray}
&&[\Omega _{{\bf p}}+\Omega _{{\bf k}-{\bf p}}-\lambda _\alpha (K)]x_{1,%
\frac{1}{2}}^{(1,\alpha )}({\bf k},{\bf p})  \nonumber \\
&&+\frac 14\sum _{{\bf p} ^{\prime }}w_{\frac{1}{2},\frac{1}{2};\frac{1}{2},%
\frac{1}{2}} \left( {\bf p},{\bf k} -{\bf p};{\bf k}-{\bf p}^{\prime },{\bf p%
}^{\prime }\right) x_{1,\frac {1}{2}}^{(1,\alpha )}({\bf k},{\bf p}^{\prime
})=0.  \label{t1}
\end{eqnarray}
This equation is apparently symmetrical in the spin indices and its solution
should describe triplet states. We marked it by setting $\kappa=1$.

The case $\sigma =0$ is more complex, because of the contributions from
triplet and singlet states. Accordingly, one obtains the system of two
equations: 
\begin{eqnarray*}
&&[ \Omega _{{\bf p}}+\Omega _{{\bf k}-{\bf p}}-\lambda _\alpha
(K)]x_{0,\frac 12}^{(\kappa ,\alpha )}({\bf k},{\bf p})+  \nonumber \\
&&\frac{1}{4}\sum _{{\bf p}^{\prime }}[w_{\frac{1}{2},-\frac{1}{2};-\frac{1}{%
2},\frac{1}{2}} \left( {\bf p},{\bf k}-{\bf p}; {\bf k}-{\bf p}^{\prime },%
{\bf p}^{\prime }\right) x_{0,\frac {1}{2}}^{(\kappa ,\alpha )}({\bf k},{\bf %
p}^{\prime })+  \nonumber \\
&&w_{\frac{1}{2},-\frac{1}{2};\frac{1}{2},-\frac{1}{2}}\left( {\bf p},{\bf k}%
-{\bf p};{\bf k}-{\bf p}^{\prime },{\bf p}^{\prime }\right) x_{0,-\frac{1}{2}%
}^{(\kappa ,\alpha )}({\bf k},{\bf p}^{\prime })]=0,
\end{eqnarray*}
\begin{eqnarray}
&&[ \Omega _{{\bf p}} +\Omega _{{\bf k}-{\bf p}}- \lambda _{\alpha} (K)]
x_{0,-\frac{1}{2}}^{(\kappa ,\alpha )} ({\bf k},{\bf p})+  \nonumber \\
&&\frac{1}{4}\sum _{{\bf p} ^{\prime }}[w_{-\frac{1}{2},\frac{1}{2};-\frac{1%
}{2},\frac{1}{2}} \left( {\bf p},{\bf k}-{\bf p}; {\bf k}-{\bf p}^{\prime },%
{\bf p}^{\prime }\right) x_{0,\frac {1}{2}} ^{(\kappa ,\alpha )}({\bf k},%
{\bf p}^{\prime })+  \nonumber \\
&&w_{-\frac{1}{2},\frac{1}{2};\frac{1}{2},-\frac{1}{2}}\left( {\bf p},{\bf k}%
-{\bf p}; {\bf k}-{\bf p}^{\prime },{\bf p} ^{\prime }\right) x_{0,- \frac{1%
}{2}}^{(\kappa,\alpha )}({\bf k},{\bf p} ^{\prime })]=0.  \label{sigma0}
\end{eqnarray}
A nontrivial solution to this system exists only if both equations are
equivalent. This equivalence is guaranteed by antisymmetrical properties of
the $x$ and $w$ functions: 
\[
x_{0,\frac {1}{2}}^{(\kappa ,\alpha )}({\bf k},{\bf p})=-x_{0,-\frac{1}{2}%
}^{(\kappa ,\alpha )}({\bf k},{\bf k-p});
\]
\[
w_{s,s^{\prime };s^{\prime \prime },s^{\prime \prime \prime }}\left( {\bf p},%
{\bf p}^{\prime };{\bf p}^{\prime \prime },{\bf p} ^{\prime \prime \prime
}\right) = -w_{s^{\prime },s;s^{\prime \prime },s^{\prime \prime \prime
}}\left( {\bf p}^{\prime },{\bf p};{\bf p}^{\prime \prime },{\bf p}^{\prime
\prime \prime }\right) =-w_{s,s^{\prime };s^{\prime \prime \prime
},s^{\prime \prime }}\left( {\bf p},{\bf p}^{\prime };{\bf p}^{\prime \prime
\prime },{\bf p}^{\prime \prime }\right).
\]
For this reason we consider only the first equation of the system (\ref
{sigma0}) in the following.

The equation for the symmetrical in spin indices solution $(x_{0,\frac{1}{2}%
}^{(1,\alpha )}({\bf k},{\bf p}^{\prime })=x_{0,-\frac 12}^{(1,\alpha )}(%
{\bf k},{\bf p}^{\prime })),$ has the form:

\begin{eqnarray}
&&\lbrack \Omega _{{\bf p}}+\Omega _{{\bf k}-{\bf p}}-\lambda _\alpha
(K)]x_{0,\frac 12}^{(1,\alpha )}({\bf k},{\bf p})+ \frac{1}{4}\sum _{{\bf p}%
^{\prime }}[w_{\frac{1}{2},-\frac{1}{2};-\frac{1}{2},\frac{1}{2}} \left( 
{\bf p},{\bf k}-{\bf p}; {\bf k}-{\bf p}^{\prime },{\bf p}^{\prime }\right)+
\nonumber \\
&&w_{\frac{1}{2}, -\frac{1}{2};\frac{1}{2},-\frac{1}{2}}\left( {\bf p},{\bf k%
}-{\bf p};{\bf k}- {\bf p}^{\prime },{\bf p}^{\prime }\right) ]x_{0,\frac{1}{%
2}} ^{(1,\alpha )}({\bf k},{\bf p}^{\prime })=0  \label{t2}
\end{eqnarray}

Eqs.(\ref{t1}) and (\ref{t2}) describe the triplet states with different
spin projection. It will be shown below that these equations are equivalent.

The equation for the antisymmetrical in spin indices solution to the system (%
\ref{sigma0}) $(x_{0,\frac{1}{2}}^{(0,\alpha )}({\bf k},{\bf p}^{\prime
})=-x_{0,-\frac{1}{2}}^{(0,\alpha )}({\bf k},{\bf p}^{\prime }))$ describes
the singlet states:

\begin{eqnarray}
&&[\Omega _{{\bf p}}+\Omega _{{\bf k}-{\bf p}}-\lambda _\alpha
(K)]x_{0,\frac {1}{2}}^{(0,\alpha )}({\bf k},{\bf p}) +\frac{1}{4}\sum _{%
{\bf p}^{\prime }}[w_{\frac{1}{2},-\frac{1}{2};-\frac{1}{2},\frac{1}{2}}
\left( {\bf p},{\bf k}-{\bf p};{\bf k}-{\bf p}^{\prime },{\bf p}^{\prime
}\right)-  \nonumber \\
&&w_{\frac{1}{2},-\frac{1}{2};\frac{1}{2},-\frac{1}{2}}\left( {\bf p},{\bf k}%
- {\bf p};{\bf k}-{\bf p}^{\prime },{\bf p}^{\prime }\right) ]x_{0,\frac{1}{2%
}} ^{(0,\alpha )}({\bf k},{\bf p}^{\prime })=0  \label{s2}
\end{eqnarray}

It is useful to express the interaction matrix $w$ in Eqs.(\ref{t1}),(\ref
{t2}) and (\ref{s2}) in terms of the Hamiltonian parameters. In a more
detailed form the Hamiltonian in the site representation may be written as:

\begin{eqnarray}  \label{e10}
&&H=\frac{1}{4}\sum _{mnm\prime n\prime }U_{mn,m\prime n\prime
}^{(1)}c_{m,s}^{+}c_{n,s}^{+}c_{m^{\prime },s}c_{n^{\prime },s} \\
&&+\frac{1}{4}\sum _{mnm^{\prime }n^{\prime }}U_{mn,m^{\prime }n^{\prime
}}^{(2)}(c_{m,s}^{+}c_{n,-s}^{+}+c_{m,-s}^{+}c_{n,s}^{+})(c_{m^{\prime
},s}c_{n^{\prime },-s}+c_{m^{\prime },-s}c_{n^{\prime },s})  \nonumber \\
&&-\frac 14\sum _{mnm^{\prime }n^{\prime }}U_{mn,m^{\prime }n^{\prime
}}^{(3)}(c_{m,s}^{+}c_{n,-s}^{+}-c_{m,-s}^{+}c_{n,s}^{+})(c_{m^{\prime
},s}c_{n^{\prime },-s}-c_{m^{\prime },-s}c_{n^{\prime },s}),\ s=\frac{1}{2} .
\nonumber
\end{eqnarray}
The first two terms are responsible for the triplet interaction (matrices $%
U_{mn,m^{\prime }n^{\prime }}^{(1)},U_{mn,m^{\prime }n^{\prime }}^{(2)}$ are
antisymmetrical) whereas the third term describes the singlet interaction
(matrix $U_{mn,m^{\prime }n^{\prime }}^{(3)}$ is symmetrical). A comparison
of the two equivalent expressions for the Hamiltonian (\ref{e1}), (\ref{e10}%
) gives:

\begin{eqnarray}
&&U_{m,s;n,s;m^{\prime },s;n^{\prime },s}=U_{m,-s;n,- s;m^{\prime
},-s;n^{\prime },-s}=U_{mn,m^{\prime }n^{\prime }} ^{(1)};  \nonumber \\
&&U_{m,s;n,-s;m^{\prime },s;n^{\prime },-s}=U_{m, s;n,s;m^{\prime
},-s;n^{\prime },s}=U_{mn,m\prime n\prime }^{(2)}-U_{mn,m\prime n\prime }
^{(3)};  \nonumber \\
&&U_{m,s;n,-s;m^{\prime },-s;n^{\prime },s}=U_{m,-s;n,s;m^{\prime},s;
n^{\prime },-s}=U_{mn,m^{\prime }n^{\prime}} ^{(2)}+ U_{mn,m^{\prime
}n^{\prime }} ^{(3)}.  \label{e111}
\end{eqnarray}
Here the site and spin indices are shown separately for the sake of clarity.
It seems reasonable to suppose that the form of the interaction depends only
on the spin of a 4-fermion state, rather than on its spin projection. It
leads to an equality $U_{mn,m^{\prime }n^{\prime }}^{(1)}=U_{mn,m^{\prime
}n^{\prime }}^{(2)}$ , that, being not very important, makes calculations
more easy. Using Eqs.(\ref{n1}), (\ref{n2}) in the momentum representation
we may insert the Hamiltonian parameters (i.e. matrixes $U^{\left( 1\right)
},U^{\left( 2\right) }$ and $U^{\left( 3\right) }$) into Eqs.(\ref{t1}),(\ref
{t2}) and (\ref{s2}) instead of the function $w$. After some transformations
one obtains the equation for the singlet states: 
\begin{eqnarray}  \label{e12}
&&[\Omega _{{\bf p}}+\Omega _{{\bf k}-{\bf p}}-\lambda _{0,\alpha }({\bf k}%
)]x_{0,s}^{(0,\alpha )}({\bf k},{\bf p})+ \\
&&\frac{1}{2}(th\frac{\beta \Omega _{{\bf p}}}{2}+th\frac{\beta \Omega _{%
{\bf k}{\bf p}}}{2})\sum _{{\bf p}^{\prime }}U_{{\bf p},{\bf k}-{\bf p}; 
{\bf k}-{\bf p}^{\prime },{\bf p}^{\prime }} ^{(3)}x_{0,s}^{(0,\alpha )} (%
{\bf k},{\bf p}^{\prime })=0,\ s=\pm \frac{1}{2}  \nonumber
\end{eqnarray}
Both of equivalent Eqs.(\ref{t1}),(\ref{t2}) for the triplet states may be
transformed to: 
\begin{eqnarray}  \label{e13}
&&[\Omega _{{\bf p}}+\Omega _{{\bf k}-{\bf p}}-\lambda _{1,\alpha }({\bf k}%
)]x_{\sigma ,s}^{(1,\alpha )}({\bf k},{\bf p})+ \\
&&\frac{1}{2}(th\frac{\beta \Omega _{{\bf p}}}2+th\frac{\beta \Omega _{{\bf k%
} {\bf p}}}{2})\sum _{{\bf p}^{\prime }}U_{{\bf p},{\bf k}-{\bf p};{\bf k}- 
{\bf p}^{\prime },{\bf p}^{\prime }} ^{(1)}x_{\sigma ,s}^{(1,\alpha )}({\bf k%
},{\bf p}^{\prime })=0,\ \sigma =0,\pm 1.  \nonumber
\end{eqnarray}
Equations (\ref{e12}),(\ref{e13}) may be solved numerically though they are
too complicated yet. As it will be shown below, the account of the finite
number of sites in the interaction parameters $U^{(1)}, U^{(2)}$, $U^{(3)}$
permits to simplify the calculations significantly. An example of such a
model is discussed in the next section.

\bigskip

\subsection{\bf Spectra of 2-particle excitations. 1 -- 2 model}

\bigskip

Let us fix up the form of the fermion -- fermion interaction. It takes into
account the interaction on the same or neighboring sites (1 -- 2 model). The
interaction parameters $U^{(1)},U^{(2)},U^{(3)}$ (see expression (\ref{e10}%
)) in the momentum representation have the form:

\begin{eqnarray*}
&U_{{\bf p},{\bf p}-{\bf k};{\bf k}-{\bf p}^{\prime },{\bf p}^{\prime
}}^{(1)}=U_{{\bf p},{\bf p}-{\bf k};{\bf k}-{\bf p}^{\prime },{\bf p}
^{\prime }}^{(2)}=\delta _{{\bf k}{\bf k}^{\prime }}U_{t2}\sum _{{\bf l}%
}[cos({\bf p}-{\bf p}^{\prime },{\bf l)}-cos({\bf k}-{\bf p}-{\bf p}^{\prime
},{\bf l)}] \\
&U_{{\bf p},{\bf p}-{\bf k};{\bf k}-{\bf p}^{\prime },{\bf p}^{\prime
}}^{(3)}=\delta _{{\bf k}{\bf k}^{\prime }}\{U_{s1}+U_{s2}\sum _{{\bf l}%
}[cos({\bf p}-{\bf p}^{\prime },{\bf l)}+cos({\bf k}-{\bf p}-{\bf p}^{\prime
},{\bf l)}]\},
\end{eqnarray*}
where $U_{t2}$ is the interaction energy of the fermions in the triplet
state and $U_{s2},U_{s1}$ are the intersite and onsite interaction of the
fermions in the singlet state; ${\bf l}$ is a lattice vector. Substituting
these expressions into Eqs.(\ref{e12}, \ref{e13}) one can obtain: 
\begin{eqnarray}  \label{add0}
x_{\sigma ,s}^{(1,\alpha )}({\bf k},{\bf p})- 2L_{kp}U_{t2}\sum _{{\bf l}%
}sin(\frac{1}{2}{\bf k}-{\bf p,l)}\sum _{{\bf p}^{\prime }}sin(\frac{1}{2} 
{\bf k}-{\bf p}^{\prime },{\bf l)} x_{\sigma ,s}^{(1,\alpha )}({\bf k}, {\bf %
p}^{\prime })=0
\end{eqnarray}
- for the triplet state and

\begin{eqnarray}
x_{0,s}^{(0,\alpha )}({\bf k},{\bf p})-L_{kp}\sum _{{\bf p}^{\prime
}}[U_{s1} +2U_{s2}\sum _{{\bf l}}cos(\frac{1}{2}{\bf k}-{\bf p,l)}cos(\frac{1%
}{2}{\bf k}-{\bf p}^{\prime },{\bf l)}] x_{0,s}^{(0,\alpha )}({\bf k},{\bf p}%
^{\prime })=0  \label{add1}
\end{eqnarray}
-for the singlet one. Here the notations are: 
\begin{eqnarray*}
L_{kp}=\frac{1}{2}\frac{th(\beta \Omega _{{\bf p}}/2)+th(\beta \Omega _{{\bf %
k}-{\bf p}}/2)}{\lambda _{(\kappa ,\alpha )}({\bf k})- \Omega _{{\bf p}%
}-\Omega _{{\bf k}-{\bf p}}},\ \Omega _{{\bf k}}=\omega _{{\bf k}}-\mu
=t\sum _{{\bf l}}cos({\bf k}{\bf l})-\mu ,
\end{eqnarray*}

${\bf p}$ is the internal momentum variable and the hopping integral is
supposed to be $t({\bf m} -{\bf n} )=t\neq 0$ only if ${\bf m}={\bf n} \pm 
{\bf l}$. These equations obviously lead to the following ${\bf p}$ --
dependence of the pair wave function: 
\begin{eqnarray}  \label{e14}
x_{\sigma ,s}^{\left( 1,\alpha \right) }({\bf k,p)=} L_{kp}\sum%
\limits_{l=1}^{dim}A_l^{\left( t\right) }({\bf k})\sin (\frac 12 {\bf k}-%
{\bf p,l)},
\end{eqnarray}

\begin{eqnarray}  \label{e15}
x_{0,s}^{\left( 0,\alpha \right) }({\bf k,p)=}L_{kp}[A_0^{\left( s\right) }(%
{\bf k})+\sum\limits_{l=1}^{dim}A_l^{\left( s\right) }({\bf k} )\cos (\frac
12{\bf k}-{\bf p,l)}],
\end{eqnarray}
where index $l=1$ for ${\bf l} =(l_x ,0,0)$, $l=2$ for ${\bf l} =(0,l_y ,0)$%
, $l=3$ for ${\bf l} =(0,0,l_z )$ and $dim$ stands for the dimensionality of
the system. The lattice vector ${\bf l}$ numerates in a natural way the
triplet bands and therefore the number of the triplet bands is equal to the
space dimensionality. The coefficients $A_0^{\left( s\right) }({\bf k}%
),A_l^{\left( s\right) }({\bf k}) $ $A_l^{\left( t\right) }({\bf k})$ are
determined by substitution of these functional forms into Eqs.(\ref{add0}, 
\ref{add1}) for the wave functions. The equation describing the triplet
pairs appears to be diagonal after this substitution and is transformed into
linear independent algebraic equations for the functions $A_l^{\left(
t\right) }({\bf k})$: 
\begin{eqnarray*}
[1-2U_{t2}H_l({\bf k})]A_l^{\left( t\right) }({\bf k})=0,l \leq dim
\end{eqnarray*}
where $H_l=\sum _pL_{kp}\sin {}^2(\frac{1}{2}{\bf k}-{\bf p},{\bf l})$.

The excitation spectra are obtained from the equations: 
\begin{eqnarray}  \label{e16}
1-2U_{t2}H_l({\bf k})=0,
\end{eqnarray}
while the coefficients $A_l^{\left( t\right) }({\bf k})$ are derived from
the normalization condition (\ref{norm.x}) which is now written as $\sum
_{p,s}x_{\sigma ,s}^{\left( 1,\alpha \right) *}({\bf k,p)}x_{\sigma
,s}^{\left( 1,\alpha ^{\prime }\right) }({\bf k,p)}=\delta _{\alpha ,\alpha
^{\prime }}$. With the account of Eq.(\ref{e14}) one obtains: 
\begin{eqnarray*}
A_l^{\left( t\right) }=\left( \sum _pL_{kp}^2\sin ^2(\frac{1}{2}{\bf k}-{\bf %
p,l)}\right) ^{-1/2}.
\end{eqnarray*}
As a result of the diagonalisation the triplet wave functions (\ref{e14})
appear to be rather simple: 
\begin{eqnarray}  \label{e14prim}
x_{\sigma ,s}^{\left( 1,\alpha \right) }({\bf k,p)=} L_{k,p}A_l^{\left(
t\right) }({\bf k})\sin (\frac{1}{2}{\bf k}-{\bf p,l)}
\end{eqnarray}
The singlet pair case is more complicated. Nevertheless one can obtain a
linear algebraic system of 4 equations on the coefficients $A_0^{\left(
s\right) }({\bf k}),A_l^{\left( s\right) }({\bf k})$ by substituting
expression (\ref{e15}) into Eq.(\ref{add1}): 
\begin{eqnarray}  \label{e17}
(\hat 1-\hat M)\left( 
\begin{array}{c}
A_0 \\ 
A_1 \\ 
A_2 \\ 
A_3
\end{array}
\right) =0,
\end{eqnarray}
where $\hat M$ is the matrix of rank 4:

\begin{eqnarray}  \label{e17a}
\hat M=\left( 
\begin{array}{cccc}
U_{s1}I_0 & U_{s1}I_1 & U_{s1}I_2 & U_{s1}I_3 \\ 
2U_{s2}I_1 & 2U_{s2}I_{11} & 2U_{s2}I_{12} & 2U_{s2}I_{13} \\ 
2U_{s2}I_2 & 2U_{s2}I_{21} & 2U_{s2}I_{22} & 2U_{s2}I_{23} \\ 
2U_{s2}I_3 & 2U_{s2}I_{31} & 2U_{s2}I_{32} & 2U_{s2}I_{33}.
\end{array}
\right)
\end{eqnarray}
and 
\begin{eqnarray*}
&I_0=\sum _pL_{kp};\ I_l=\sum _pL_{kp}\cos (\frac{1}{2}{\bf k} -{\bf p,l)};
\\
&I_{lm}=\sum _pL_{kp}\cos (\frac{1}{2}{\bf k}-{\bf p},{\bf l})\cdot \cos (%
\frac{1}{2}{\bf k}-{\bf p,m)}.
\end{eqnarray*}
The matrix $\hat M$ is written for the 3 -- dimensional system. The same
matrix for the 2 -- dimensional system may be obtained by omitting the last
column and the last line. For a 1 -- dimensional system this simple
procedure must be repeated. The spectra of the singlet states are found by
equating the determinant of the system (\ref{e17}) to zero: 
\begin{eqnarray}  \label{e18}
Det(\hat 1-\hat M)=0,
\end{eqnarray}
whereas the structure coefficients $A_0^{\left( s\right) }({\bf k}%
),A_l^{\left( s\right) }({\bf k})$ describing the wave function of the
singlet pair (\ref{e15}) are found from Eq.(\ref{e17}) and the normalization
condition:

\begin{eqnarray*}
\sum _{p,s}x_{0,s}^{\left( 0,\alpha \right) }({\bf k,p)}^{*}x_{0,s}^{\left(
0,\alpha ^{\prime }\right) }({\bf k,p)}=\delta _{\alpha ,\alpha ^{\prime }}.
\end{eqnarray*}

Now we have to find the excitation spectra of the local pairs by utilizing
formulae (\ref{e16}) -- (\ref{e18}). The numerical calculations have been
done in the units: $D_{pol}=4$. The results are represented in Fig.1 for the
interaction value of $U_{s1}=+8$ (repulsive onsite interaction term); $%
U_{s2}=-6.8,U_{t2}=-6.5$ (attractive intersite terms for both the singlet
and triplet states).

Here the most interesting case is considered: onsite repulsion (Coulomb
force predominance) and intersite attraction. The concrete value of onsite
repulsion has only a small effect on the character of band structure. The
explanation is clear: two coupled fermions are placed on neighbor sites
mostly where attraction dominates and "don't feel" onsite repulsion. On the
contrary, the value of the intersite attraction is important for the
determination of low -- lying bands. This situation has been analyzed in
detail in \cite{Sofo}. The authors argued that the gap between the local
pair band and the polaron band, that is placed above, increases with the
coupling constant when considering onsite attractive Hubbard model. The
substitution of the onsite attraction by the intersite one does not affect
this qualitative result. Fig.1 represents band structure of the singlet and
triplet pairs at definite values of s-wave and p-wave attraction. The choice
of the attraction value $U_{s2}=-6.8$ and $U_{t2}=-6.5$ gives reasonable
mass of local pairs.

It becomes evident that the band structure is not as simple as in the
standard 1-site attractive Hubbard model. There are additional gaps between
the singlet and triplet bands, along with large ordinary gap. Each of the
singlet and triplet band in its turn is splitted into two subbands . This
could explain those contradictions that seem take place in determination of
the gap value from different experiments. A one -- particle (polaron) band
is lying higher with the energy of its bottom being equal $0.97$. It is
necessary now to normalize our relative units. An appropriate value of a
polaron bandwidth is about 50-100meV. The undimentional polaron bandwidth is
equal 4 in the units that were used throughout the paper. It means that our
energy unit is about $20meV$, The chosen value of the intersite attraction
is $136meV(6.8\cdot 20meV)$, that seems reasonable. The value of the gap
between the lowest local pair energy and lowest polaron energy depends
strictly on the sort of pairing. Thus, in the case presented in Fg.1, the
gapwidth between polaron band and s-pair band is about $20meV(0.97\cdot
20meV)$

Another interesting conclusion concerns a mass of the local pairs. According
to the wide-spread point of view the local pair mass should be large enough
this being considered as a shortcoming of local pair theories. Our
calculations demonstrate (see Fig.1), that the situation is more favorable:
the width of the lowest singlet band is only 3-4 times smaller, than the
width of the fermion band. It gives the mass of the local pair $m^{**}\simeq
3\div 4m^{*}$, where $m^{*}$ is the effective fermion mass. One can notice
finally, that the triplet pairs appear to be several times heavier than the
singlet ones for this set of parameters.

\bigskip

\section{\bf Local pairs in external fields}

\bigskip

The excitation spectra and the local pair structure studied above may be
used for theoretical investigation of thermodynamic and kinetic properties
of the systems with the strongly interacting electrons. On the other hand
there are many experiments dealing with the interaction of HTSC materials
with different external fields. For the theoretical explanation of these
experiments one has to know the form of the interaction of the pairs with
the external fields. The corresponding vertex functions should be obviously
dependent on the internal structure of the coupled state. The interaction of
the fermion with the external field presented in Eq.(\ref{e1}) for the
Hamiltonian is of rather general form. It includes, for example, the
interaction with neutrons and electromagnetic field. By integrating over the
grassman variables in the same way as it was done before, one can obtain the
action in the form (\ref{e4}), with the matrix $Y$ depending on the external
field: 
\begin{eqnarray}  \label{matrY}
Y_{PP^{\prime }}=\left( 
\begin{array}{cc}
(r_{PP^{\prime }}+i\epsilon \delta _{PP^{\prime }})\delta _{\epsilon
\epsilon ^{^{\prime }}}+B(\epsilon -\epsilon ^{^{\prime }})_{PP^{\prime }} & 
\psi \left( \epsilon +\epsilon ^{^{\prime }}\right) _{PP^{\prime }} \\ 
\psi ^{*}\left( \epsilon +\epsilon ^{^{\prime }}\right) _{PP^{\prime }} & 
-(r_{PP^{\prime }}+i\epsilon \delta _{PP^{\prime }})\delta _{\epsilon
\epsilon ^{^{\prime }}}-B(\epsilon -\epsilon ^{^{\prime }})_{PP^{\prime }}
\end{array}
\right).
\end{eqnarray}
In Eq.(\ref{matrY}) all matrices are written in the momentum representation
and an imaginary ''time'' Fourier -- transformation is assumed. Expanding
this expression up to the second order in the fields $\psi ,\psi ^{*}$ and
to the first order in the (weak) field $B$, one can obtain the effective
action for the local pairs in the external field:

\begin{eqnarray*}
&Z=\int DC^{*}DC\exp S(C^{*},C), \\
\\
&S(C^{*},C)=C _\sigma ^{\left( \kappa \alpha \right) *}\left( {\bf k,}%
E\right)\left( {i} E-\lambda _{\kappa \alpha }({\bf k}) \right) C _\sigma
^{\left( \kappa \alpha \right) } \left( {\bf k,}E\right) \\
\\
&+ C _\sigma ^{\left( \kappa \alpha \right) *} \left( {\bf k,}E\right)
\Gamma _{\sigma \sigma ^{\prime }}^{\left( \kappa \alpha ,\kappa ^{\prime
}\alpha ^{\prime }\right) } \left( {\bf k,}E,{\bf k^{\prime },}E^{\prime }
\right) C _{\sigma ^{\prime }}^{\left( \kappa ^{\prime }\alpha ^{\prime
}\right) } \left( {{\bf k}^{\prime },}E^{\prime}\right) .
\end{eqnarray*}
Here the summation on the repeated indices is assumed. Eq.(\ref{e9}) and the
orthonormal basis (\ref{setx}) were used in the derivation of this equation.
The vertex $\Gamma $ has the form: 
\begin{eqnarray}  \label{Gamma}
\Gamma _{\sigma \sigma ^{\prime }}^{\left( \kappa \alpha ,\kappa ^{\prime
}\alpha ^{\prime }\right) }\left( {\bf k,}E,{\bf k^{\prime },}E^{\prime
}\right) =2\sum _s B_{\sigma ^{\prime }- s,\sigma -s}({\bf k-k^{^{\prime }},}%
E-E^{\prime })F_{\sigma \sigma ^{\prime }s}^{\left( \kappa \alpha ,\kappa
^{\prime }\alpha ^{\prime }\right) }({\bf k-k^{^{\prime }}}).
\end{eqnarray}
Here the formfactor 
\begin{eqnarray}  \label{formfac}
F_{\sigma \sigma ^{\prime }s}^{\left( \kappa \alpha ,\kappa ^{\prime }\alpha
^{\prime }\right) }({\bf k-k^{^{\prime }}})=\sum _{{\bf p} }x_{\sigma
,\sigma -s}^{\left( \kappa ,\alpha \right) ^{*}}({\bf k,k-p} )x_{\sigma
^{\prime },\sigma ^{\prime }- s}^{\left( \kappa ^{\prime },\alpha ^{\prime
}\right) }({\bf k^{^{\prime }},k^{^{\prime }}-p)},
\end{eqnarray}
depends only on the structure of the coupled state. The external field is
written in the more convenient form -- with the spin index extracted and
translational invariance taken into account: $B_{PP^{\prime }}(E) \equiv
B_{s_p,s_{p^{\prime }}}({\bf p}-{\bf p}^{\prime },E)$. As it was mentioned
above, the vertex of the interaction of pairs with the external field
depends on the frequency, the wave vector of the external field and the wave
functions of the fermion pairs. One can see that the slowly varying field is
not sensitive to the pair structure. A simple test shows that in this case
the vertex of the singlet local pair interaction with the external field
equals to zero. For the triplet pairs it is twice the interaction of the
fermion with the field. The behavior of formfactor (\ref{formfac}) plays a
significant role in the theoretical explanation of experiments on the
scattering of particles on the local pairs. For example, at low temperatures
and small transferred momenta the transitions appear to be suppressed. The
reason is quite obvious: at the low temperature the local pairs being bosons
are condensed at the lowest level with the momenta ${\bf k}=0$. The
transitions into the upper bands with approximately the same momentum are
suppressed due to the orthogonality of the wave functions with different
energies: $F({\bf k-k^{\prime }}=0)=0$.

The typical formfactors of the transitions from the lowest singlet band to
the upper singlet or triplet one are presented in Fig.2. There is also the
formfactor describing the transitions inside the lowest singlet band. The
transitions into the second triplet band are forbidden for chosen
transferred momenta ${\bf k}=(k_x,0)$ due to the orthogonality of the wave
functions. The evident momentum dependence of the formfactor confirms its
importance in the interpretation of experiments.

\section{Conclusion}

The mathematics of the local pair theory was developed. The {\it ad hoc}
calculation of band structure of the local pairs and their interaction with
external fields are presented here. It is shown that numerical calculations
based on analytical formulas and being relatively simple lead to exact
qualitative predictions. The method can be easily generalized. The momentum
dependency of the vertex is essential for theoretical explanation of
experiments .

\section{\bf Acknowledgments}

The author is grateful to Profs. A.S.Alexandrov, A.B.Khmelinin, A.B.Krebs,
and D.A.Samarchenko for the enlightening discussions.

This work has been supported in part by RFFI Grant No.97-02-16705.


\begin{thebibliography}{99}
\bibitem{Little}  {W.A.Little}, Phys.Rev. A134, 1416 (1964). {V.L.Ginzburg},
Contemp.Phys. 9, 355 (1968).

\bibitem{Frohlich}  {H.Frohlich}, Phys.Lett. 26A, 169 (1968).

\bibitem{Schri}  {J.R. Schrieffer et al.} Phys.Rev. {\bf B39}, 11663 (1989).

\bibitem{Krebs}  {A.S.Alexandrov, A.B.Krebs}, Uspehi.Fiz.Nauk {\bf 162}-5, 1
(1992),

\bibitem{Alex}  {A.S. Alexandrov and Sir Nevill Mott}// {\em High
Temperature Superconductors and other Superfluids} (Taylor\& Francis, 1994),
A.S. Alexandrov, N.F. Mott, Supercond. Sci. Technol. {\bf 6}, 215 (1993).

\bibitem{Sofo}  J.O.Sofo et al., Phys.Rev. B45, 9860 (1992).

\bibitem{drechsler}  M.Drechsler et al. Ann.Phys. 1, 15 (1993).

\bibitem{micnas}  {R. Micnas, J. Ranninger and S. Robaszkiewicz}, Rev. Mod.
Phys. {\bf 62}, 113 (1990),

\bibitem{boer}  {Jan de Boer, V.E.Korepin, A.Schadschneider}, Phys. Rev.
Lett. 74, 789 (1995).

\bibitem{Verdozzi}  {C.Verdozzi, M.Cini}, Phys.Rev. {\bf B51}, 7412 (1995).

\bibitem{b4}  R.P.Feynman Statistical Mechanics, California, Inst. of Techn.
(1972).

\bibitem{b2}  V.N. Popov, Path integrals in quantum field theory and
statistical physics, Moscow (1976).

\bibitem{cross}  {\ C.A.R. S\'{a} de Melo}\ {\it et al.\/}, Phys. Rev. Lett. 
{\bf 71}, 3202 (1993),

\bibitem{polaron}  {A.B.Krebs, S.G.Rubin}, Phys.Rev. {\bf B49}, 11808
(1994); {P. Sethna}, Phys.Rev. {\bf B25}, 5050 (1982).

\bibitem{RubMel}  {G.G.Melkonian, S.G.Rubin } Phys.Rev.{\bf B57},10867 (1998)

\bibitem{RubTer}  {S.G.Rubin, A.G.Terekidi}, Preprint N 024-97,MEPI, Moscow;

\bibitem{HS}  R.L.Stratonovich, DAN USSR 115, 1097 (1957), J.Habbard,
Phys.Rev. Lett. 3, 77 (1959).

\bibitem{Svidz}  {A.V. Svidzinski}, {\sl Space-inhomogeniens problems in a }

{\sl theory of superconductivity\/}, Nauka, Moskow (1982).

\bibitem{RubKon}  R.V.Konoplich,S.G.Rubin, Univ.di Roma,IFNF,preprint No
1004 (1994),

\bibitem{b1}  {A.S.Alexandrov, S.G.Rubin}, Phys.Rev. B47, 5141 (1993),

\bibitem{zetp}  {A.A.Gorbatsevich, I.V.Tocatli}, Sov.Phys. JETP {\bf 103},
702 (1993).

\bibitem{triplet}  {A.S.Alexandrov, N.F.Mott}, Int.J.Mod.Phys.{\bf 8}, 2075,
(1994).
\end{thebibliography}
\end{document}